\documentclass{ieeeaccess}

\renewcommand{\arraystretch}{1.4}

\usepackage{comment}
\usepackage{epsfig}
\usepackage{subfigure}
\usepackage{amsfonts}
\usepackage{amsmath}
\usepackage{amssymb}
\usepackage{graphicx}
\usepackage{url}
\usepackage{epstopdf}
\usepackage{graphicx}
\usepackage{array}
\graphicspath{ {./images/} }
\usepackage{cite}
\usepackage{amsmath,amssymb,amsfonts}
\usepackage{algorithmic}
\usepackage{graphicx}
\usepackage{textcomp}
\usepackage{caption,setspace}
\def\BibTeX{{\rm B\kern-.05em{\sc i\kern-.025em b}\kern-.08em
    T\kern-.1667em\lower.7ex\hbox{E}\kern-.125emX}}
\begin{document}
\history{Date of publication September 17, 2018, date of current version October 12, 2018.}
\doi{10.1109/ACCESS.2017.DOI}

\title{Decentralized Applications: The Blockchain-Empowered Software System}
\author{
\uppercase{Wei Cai}\authorrefmark{1,2}, \IEEEmembership{Member, IEEE},
\uppercase{Zehua Wang}\authorrefmark{2,3}, \IEEEmembership{Member, IEEE},
\uppercase{Jason B. Ernst}\authorrefmark{3}, \IEEEmembership{Member, IEEE},
\uppercase{Zhen Hong}\authorrefmark{2}, \IEEEmembership{Student Member, IEEE},
\uppercase{Chen Feng}\authorrefmark{4}, \IEEEmembership{Member, IEEE}, and 
\uppercase{Victor C.M. Leung}\authorrefmark{2},
\IEEEmembership{Fellow, IEEE}}
\address[1]{School of Science and Engineering, The Chinese University of Hong Kong, Shenzhen, Guangdong 518172 China}
\address[2]{Department of Electrical and Computer Engineering, The University of British Columbia, Vancouver, BC V6T1Z4 Canada}
\address[3]{Left Of The Dot Media Inc., Unit \#4 - 20000 Stewart Crescent, Maple Ridge, BC, V2X9E7 Canada}
\address[4]{School of Engineering, The University of British Columbia, Okanagan, BC V1V1V7 Canada}

\tfootnote{This work is supported by funding from the Natural Sciences and Engineering Research Council of Canada.
}

\markboth
{Wei Cai \headeretal: Decentralized Applications: The Blockchain-Empowered Software System}
{Wei Cai \headeretal: Decentralized Applications: The Blockchain-Empowered Software System}

\corresp{Corresponding author: Victor C.M. Leung (e-mail: vleung@ece.ubc.ca).}

\begin{abstract}
Blockchain technology has attracted tremendous attention in both academia and capital market. However, overwhelming speculations on thousands of available cryptocurrencies and numerous initial coin offering (ICO) scams have also brought notorious debates on this emerging technology. This paper traces the development of blockchain systems to reveal the importance of decentralized applications (dApps) and the future value of blockchain. We survey the state-of-the-art dApps and discuss the direction of blockchain development to fulfill the desirable characteristics of dApps. The readers will gain an overview of dApp research and get familiar with recent developments in the blockchain.
\end{abstract}


\titlepgskip=-15pt

\maketitle

\section{Introduction} \label{sec:introduction}

By definition, a blockchain is a continuously growing chain of blocks, each of which contains a cryptographic hash of the previous block, a time-stamp, and its conveyed data \cite{Nofer2017}. Due to the existence of the cryptographic hash, the data stored in a blockchain are inherently resistant to modification: if one block of data is modified, all blocks afterward should be regenerated with new hash values. This feature of immutability is fundamental to blockchain applications. 

Maintenance of peer-to-peer (P2P) ledgers for cryptocurrencies has become the first killer application of blockchain. Thousands of cryptographic tokens, or coins, were delivered to the public market, after the huge leap in market cap of Bitcoin \cite{bitcoin}. However, due to the lack of legal regulation and auditing, a large number of scams, so-called ``air coins'', also brought bad reputations to the blockchain technology. Doubts on the value of cryptocurrencies have been raised. Warren Buffett---the famous billionaire investor---insisted that cryptocurrencies will come to a ``bad ending'', and claimed that Bitcoin is ``probably rat poison squared''. Instead of discussing cryptocurrencies, this paper surveys the state-of-the-art of blockchain technology and introduces decentralized applications (dApps), which is a novel form of the blockchain-empowered software system.

In the rest of this article, we review the classic blockchain systems in Section~\ref{sec:classic} and reveal the value of blockchain systems in Section~\ref{sec:value}. We survey the state-of-the-art dApps in Section~\ref{sec:dApp} and envision the desirable characteristics of future dApps in Section~\ref{sec:requirement}. We also discuss the considerations when selecting a blockchain implementation in Section~\ref{sec:selection}. Recent research to develop next-generation blockchain systems that address some of these characteristics is presented in Section~\ref{sec:development}. Section~\ref{sec:conclusion} concludes the article.

\section{Background: Classic Blockchain Systems}\label{sec:classic}

In this section, we trace the evolution of decentralized ledgers that led to classic blockchain systems adopting public consensus models.

\subsection{Prehistory} \label{sec:prehistory}

Researchers have been working on the implementation of digital cash \cite{untraceablepayment} since the 1980s. Before the advent of Bitcoin, academia has established solid foundations in this topic. The blockchain concept, the fundamental form of public ledger, was first introduced for time-stamped digital documents in 1991 \cite{Haber1991}. Later, Merkle tree \cite{Merkle} was incorporated into the cryptographically secured chain by allowing several documents to be collected into one block, which improves the system efficiency and reliability \cite{merklechain}. However, such a ledger implemented with a chain of blocks is still a centralized database, which relies on the maintenance of a trusted third party financial institute. 

\subsection{Synchronization Issue}

Centralized systems are criticized for their vulnerability, due to the single-point-of-failure (SPOF) issue. By contrast, decentralized systems implemented in a distributed manner suffer from the data synchronization issue, which is well summarized as the Byzantine Generals' Problem \cite{Byzantine}. In other words, the participants in the decentralized ledger system need to achieve consensus, an agreement upon every message being broadcast to each other. A common Byzantine fault tolerance can be achieved if the ``loyal generals'', the honest peers in our context, have a majority agreement on their decisions. Nevertheless, intruders may perform Sybil attack \cite{Sybil2002} to control a substantial fraction of the public P2P system by representing multiple identities, which may lead to a critical ``Double Spending'' issue in the blockchain-empowered decentralized ledger.

\subsection{Double Spending Issue}

Thanks to the hash-linking feature of the blockchain, each coin in the ledger can be traced back to the first record when it was created. Therefore, forgery on a non-existing coin is impossible in a public decentralized ledger. However, different from a physical coin, a digital coin can be easily replicated by duplicating the data. In this context, it is critical to prevent the dishonest behavior of spending a coin more than once. If a dishonest user of the public ledger is capable of performing a Sybil attack, the coins that the user double-spends will be legalized by the majority of parties, which diminishes user trust as well as the circulation and retention of the currency. 

\subsection{Proof of Work Consensus} \label{sec:pow}

Satoshi Nakamoto applied Proof-of-Work (PoW) \cite{Hashcash2002} to solve the double spending issue in the first white paper of Bitcoin \cite{bitcoin}. In this case, the PoW involves a mathematical calculation to scan for a numeric value that when hashed, the hash result begins with a specific number of zero bits. With PoW, each peer in the P2P network needs to compete with each other in solving the puzzles, which is also called mining. The winner of each competition will have the privilege to create a block and broadcast it to the peers. This PoW is intrinsically a brute-force search procedure, while its answer can be easily verified with a hashing process that requires $O(1)$ complexity. The PoW imposes an intentional computational cost that increases the difficulty of the identity forge in Sybil attack to a very high level, due to the large hardware investment required of a particular network participant. On the other hand, the peers who successfully create some blocks will receive coin rewards for their work. In fact, even if a particular peer has a tremendous computational capacity, the value of using this capacity to earn coin rewards is higher than that of attacking the decentralized system. This type of PoW consensus mechanism demotivates the intruders, and thus protects the decentralized ledger.   

\subsection{Broader Definition of Blockchain Systems}

As discussed above, the conventional definition of ``blockchain'' goes beyond the technology of blockchain that links data blocks into an immutable chain. It is applicable to a completely distributed and decentralized system that requires all participating peers to follow specific blockchain rules in achieving data synchronization.  In this article, we would like to present a broader definition for blockchain systems, which is a combination of the blockchain, P2P network, and the consensus model.

\begin{figure}[htp]
\centering
\includegraphics[width=85mm]{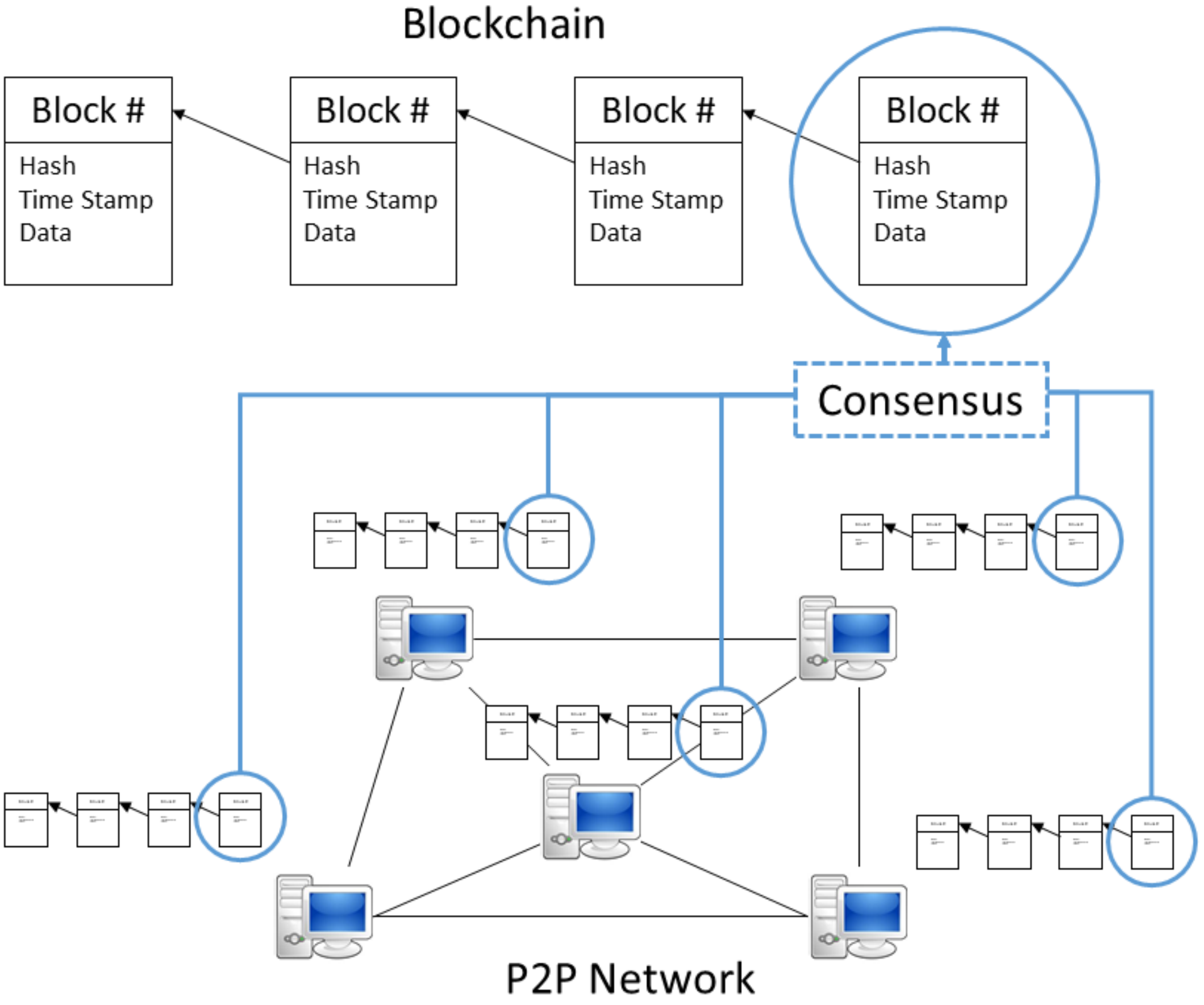}
\caption{Key Elements of Blockchain Systems}
\label{fig:blockchain}
\end{figure}

Figure \ref{fig:blockchain} illustrates the architecture of such a broadly defined blockchain system. All participants in the P2P network need to store blockchain data on their own while synchronizing all of their blocks with those stored by other peers based on a consensus model. In fact, the consensus is represented by the longest chain agreed upon by the majority of the peer nodes.

\section{Evolution of Blockchain Systems}\label{sec:value}

In this section, we discuss the evolution of different generations of blockchain systems in terms of their functions and applications.

\subsection{Decentralized Ledger}

Bitcoin \cite{bitcoin} is representative of the classic blockchain system. As the first decentralized ledger, it has attracted more than ten thousand nodes to establish the largest market capitalization among all cryptocurrencies. The most important contribution of Bitcoin is that it solves the double spending issue to make digital asset unique and valuable\footnote{More precisely, Bitcoin solves the double spending issue with high probability under the assumption of honest majority. See, e.g., Section~11 in the Bitcoin whitepaper \cite{bitcoin}.}.  In fact, the success of Bitcoin opened the door of blockchain applications to the public. However, Bitcoin itself is only a public decentralized ledger without any subject matter, which is criticized by many economists as another Ponzi scam. Along with the development of the P2P network, the subject matter of Bitcoin has now become the computational cost of nodes (miners), which is mainly concerned with PoW efforts. However, these efforts do not bring any value but only strengthen the robustness of the system. By convention, such application of decentralized ledger is called blockchain 1.0.

\subsection{Decentralized Smart Contract}

In order to add more values to the blockchain ecosystem, Ethereum \cite{ethereum} is designed to be a platform to facilitate decentralized smart contracts via Ether, its own currency vehicle. Smart contract \cite{smartcontractlogistics} refers to the idea that legal contracts can be notarized and executed automatically. Equipped with Solidity \cite{solidity}, a Turing-complete programming language, Ethereum developers are able to implement a series of smart contracts, which are executable programs written into blocks. Due to the immutable nature, Ethereum extends the application of blockchain from the data to the computation domain. In other words, after the developers have compiled and deployed their software to the public, nobody could ever revise the logic of the program. Therefore, publishing a smart contract creates a set of public trusted functionality for public users. These smart contracts, when invoked, will be executed by the distributed nodes in a decentralized manner. Applications of smart contracts are currently still in a preliminary stage. Most of the current applications are limited to the possession and transfer of virtual assets, such as stocks, bonds, game items, etc. For example, Initial Coin Offerings (ICOs) on Ethereum have become a popular solution for fundraising by start-up companies. By convention, Ethereum is considered the representative of blockchain 2.0 applications.

\subsection{Decentralized Applications}

Nevertheless, current blockchain-based applications are still limited to utilizing smart contract for core data and functionality that should be resistant to modifications. The smart contract users still need to run their programs locally in order to complete the application. One of the key reasons is the performance limitation of current blockchain technologies, which cannot meet the requirements of many applications. This leaves potential issues in operational security and application maintenance. For example, there might be intentional cheating behaviors in local pieces that are hidden from the public audit. 

To this end, the ultimate blockchain application should be a dApp that is completely hosted by P2P blockchain system. Ideally, a deployed dApp will need no maintenance and governance from the original developers. In other words, an ideal blockchain application or service should be operable without any human intervention, which forms a Decentralized Autonomous Organization (DAO) \cite{DAO}. A DAO is an organization that is run through rules encoded as smart contracts running on the blockchain. Due to its autonomous and automatic nature, a DAO's cost and profit are shared by all participants by simply recording all activities into the blocks. In fact, Bitcoin, the most classic blockchain system, is an example of a DAO. According to the definition of dApps in \cite{dApp}, dApps are characterized by four properties as follow: 

\begin{itemize}
    \item \emph{Open Source:} Due to the trusted nature of blockchain, dApps need to make their codes open source, so that audits from third parties become possible.
    \item \emph{Internal Cryptocurrency Support:} Internal currency is the vehicle that runs the ecosystem for a particular dApp. With tokens, it is feasible for a dApp to quantify all credits and transactions among participants of the system, including content providers and consumers. 
    \item \emph{Decentralized Consensus:} The consensus among decentralized nodes is the foundation of transparency. 
    \item \emph{No Central Point of Failure:} A fully decentralized system should have no central point of failure since all components of the applications will be hosted and executed in the blockchain.
\end{itemize}

\section{State-of-the-Art dApps}\label{sec:dApp}

Blockchain technology has been adopted in many industries. As summarized in "State of The dApps"  website\footnote{https://www.stateofthedapps.com/rankings}, Ethereum has hosted different categories of dApps, including exchange, energy, finance, health, identity, insurance, media, etc. However, many state-of-the-art dApps are in fact only partially decentralized. For example, Blockstack\footnote{https://blockstack.org/} and OpenBazaar\footnote{https://www.openbazaar.org/} are leveraging the blockchain to validate only identities of users and not anything else. In this section, we present a review of the existing dApps that are most popular.

\subsection{Games}

The video game industry perfectly fits the nature of cryptocurrencies ecosystem since it fulfills the ultimate dream of many game players: the items owned by their virtual characters in the gaming world are non-fungible and can be traded and inherited into a new game. To this end, the blockchain-based game is a new emerging trend. Currently, due to the limitations of transaction fee and delay, most blockchain games are still in preliminary stage, focusing on collectibles and trade of virtual assets. Even though this kind of game is not much fun at all, it still has brought a huge change in the game industry.

As one of the most successful blockchain games and even a milestone in the development of Ethereum, CryptoKitties\footnote{https://www.cryptokitties.co/} may be the most well-known blockchain game nowadays. Due to its popularity, its transactions once brought down the Ethereum network and put pressure on blockchain technology. In CryptoKitties, players can buy, sell, and breed cats by using a smart contract on the Ethereum Blockchain. Being different from previous collectible blockchain games that can only buy and sell specific items, this game is unique in differentiating each CryptoKitty in the game. Each cat is different from others in its physical characteristics, traits, and genes. A cat is bred by a couple and inherits facets from its parents as a unique combination of the two. Players are incentivized to breed cats with rare traits  \cite{CRYPTOKITTIES}\cite{CRYPTOKITTIES2}. Similar gaming mechanisms have been applied to different virtual assets to create many other blockchain games, such as Etheremon\footnote{https://www.etheremon.com/}, CryptoCelebrities\footnote{https://cryptocelebrities.co/}, CryptoCountries\footnote{https://cryptocountries.io/}, Etherbots\footnote{https://etherbots.io/}, etc.

Another representative type of blockchain games is the digital casino. The nature of cryptocurrency makes it extremely simple for these games to be developed and broadcasted. For example, Etheroll\footnote{https://etheroll.com/} enables players to bet on certain numbers for profit. Similar games include Vdice\footnote{http://www.vdice.io/}, bitcasino\footnote{https://bitcasino.io/}, VegasCasino\footnote{https://vegascasino.io/}, etc. Ponzi games\footnote{https://www.finder.com.au/a-brief-history-of-cryptocurrency-ponzi-games-up-to-fomo3d}, e.g. Fomo3D\footnote{https://exitscam.me/play}, also falls into this category.

Apparently, blockchain based games benefit from the features of non-fungible tokens and system transparency. It is good news for game players that blockchain has become a disrupting technology for the game industry. The relationship between game players and game companies has been completely transformed by such a new concept. In this ecosystem, the game players become parts of the game and create unique contents in the game, and their behaviors in games can unpredictably influence the development of the game. The virtual world in games becomes a real Utopia \cite{Dragonereum}.  

However, games on blockchains are still in their preliminary stage. First, the entertainment value of blockchain games is still far behind traditional video games. As discussed above, most blockchain games still stay at the level of exchanging collectibles no matter how the game designers change their trade method. A game that only collects tokens without any possibility for interaction is not able to attract many game players. Second, many game players play the games only for monetary purpose rather than for enjoyment. Users are just buying tokens with some visual representation, such as celebrity photos, stamps, and countries, hoping to trade them for profit. Last, the lifetime of games is unpredictable. In conventional gaming operation, parameters and rules for in-game economy and battles would be dynamically adjusted according to the progress of the game, in order to achieve better balance. Nevertheless, in a fully decentralized blockchain game, operators may lose control over the ecosystem, which may lead to rapid loss of game populations. 

Overall, while blockchain games have just been introduced several months ago, they have already attracted a lot of attention. Many giant game companies and great game producers have seen the potential of blockchain games and started to develop blockchain-based games. We expect to see some high-quality blockchain games in the near future.

\subsection{User-Generated Content (UGC) Network}

User-generated content (UGC), also known as user-created content (UCC), is used to describe any form of content, such as video, blogs, discussion post, that is created and published by a user for consumption by other users. In a UGC application, users and their contents are the core value of the system. Popular UGC applications include Reddit\footnote{https://www.reddit.com/}, 9GAG\footnote{https://9gag.com/}, Flickr\footnote{https://www.flickr.com/}, and Wikipedia\footnote{https://www.wikipedia.org/}. Existing UGC applications have critical issues regarding security and privacy. First, the original content from some small creator is easily stolen by other popular pages. Second, these giant social media platforms are privy to collect users' information and sell their private information to advertisers so that they can target users for advertisement. Blockchains are able to solve these problems due to their decentralized nature. Below we describe three prominent blockchain-based UGC platforms.

\subsubsection{Steem} Steem\footnote{https://steem.io/} is a blockchain-based platform with cryptocurrency rewards to publishers. Steem also has its own cryptocurrency, called STEEM. STEEM is available for purchase and exchange for various cryptocurrencies \cite{steem}.
Steem has proposed an idea of mining by human intelligence. People can convert their original creations, such as articles, music, and other forms of creation to money in this platform and no transaction fee is charged by a third party.

\subsubsection{Gems} According to the white paper, Gems\footnote{https://gem.co/} is a decentralized human task crowd-sourcing protocol on the Ethereum blockchain. Similar to Amazon Mechanical Turk (MTurk)\footnote{https://www.mturk.com/}, Gems is a marketplace where requesters publish their micro tasks and deploy workers to finish the tasks by paying the workers. However, MTurk, as a middleman, charges a large amount of money as transaction fees. In addition, as the accuracy of results from workers is variable, the requesters have to repetitively pay for the same tasks to reach a consensus. Gems is designed to solve the above problems. The Gems Protocol includes a staking mechanism to ensure task completion, a Gems Trust Score to value workers' integrity, and a payment system to reduce transaction fees \cite{GemsWhitePaper}.

\subsubsection{ONO} The goal of ONO\footnote{https://www.ono.chat/en/} is to establish a decentralized social network based on the principles of freedom, equality, and
social public governance, in which the value of attention is properly defined
and the content creators can fully reap the true rewards of the value they create. According to their white paper, the ONO platform will share the profit of social networking with the content creators.

\subsection{Internet of Things}

 Internet of Things (IoT) refers to the connection of billions of physical devices equipped with sensors and/or actuators to the Internet for collecting and sharing data and controlling our environment. The data can be collected and fused for communications without any human involvement, in order to bridge the digital and physical worlds \cite{internet_of}. Blockchain-based IoT solutions are well suited for simplifying business processes, improving customer experience and achieving significant cost efficiency \cite{IoT_and_blockchain}. According to a previous study \cite{IoT_and_blockchain2}, blockchain offers good potential for IoT solutions, because IoT applications are by definition distributed. Moreover, blockchain is designed as a basis for applications that involve transactions and interactions. 
 
\subsubsection{Smart Hardware}

Automation is a key concept in IoT applications. Smart hardware that connects to the network should be able to perform predefined actions without human intervention. This requirement perfectly fits the nature of smart contracts running on blockchains. With the transparent and immutable smart contracts, multiple parties in an IoT platform can establish trustful relationships without complicated conversations and regulations. For example, a guest checking into a future hotel may not need to register at the front desk, but instead pay for the room through a smart contract, which then instructs the door and all smart appliance in the specific room to accommodate the customer. On the other hand, the customer who has run out of funds will not be able to access the room or the facilities in it.

\subsubsection{Supply Chain}

IoT is bringing tremendous impact to supply chains. In the blockchain era, the integration of smart contracts with supply chains will further optimize the systems.  Supply chain management involves multiple stakeholders and considerable complexity. Multiple levels of suppliers, manufacturers, service providers, distributors, and retailers make record-keeping and communications inefficient. IoT and smart contracts can simplify the whole procedure by coordinating sensory data, documentation, and transparency to regulations. For example, a delay in the shipment of some raw material can be detected by the IoT network and its contingent plan specified in a transparent smart contract can be automatically executed to place make-up orders, so that the impact on the manufacturing process can be minimized. In this case, numerous emails and telephone communications are replaced by a commonly agreed smart contract, which can save a huge amount of time and resources.

\subsubsection{Source Tracing}

Nowadays, governments and consumers are increasingly demanding transparency regarding the sources of the goods that reach the marketplace. However, such transparency is difficult to achieve due to the large number of parties involved in the manufacturing, transportation and distribution of the goods and the diverse documentation and tracking systems that may exist between the sources and the consumer. Blockchains can fill the gap in enabling source tracing for items due to the fact that a blockchain can store an immutable transactions history on the chain, making it easy to recreate the history and identify the origin of a product. According to \cite{track}, even though a centralized system can achieve the same result in a fast speed, in many cases, it is hard to identify the source if e.g., the food purchased by a consumer get contaminated, since a trusted central agency usually does not exist, and even if one exists, there is a lack of transparent data storage in the central agency. Moreover, diverse information systems used by different parties have no motivation to be interoperable, i.e., people do not have the motivations or easy means to provide data directly to a central agency even if one exists.

\subsection{Sharing Economy Credits}

A sharing economy requires a credit system to encourage contributions from system participants and maintain fairness among them. However, traditional credits issued from a centralized commercial organization may not be considered a real incentive, since the value of the credit may be dictated by the organization, while the participants may need to withdraw and utilized these credits somewhere or for something else. This section discusses the possibility of leveraging blockchain for such an ecosystem.

\subsubsection{File Sharing Credits}

The possibility of file sharing has been investigated since the explosive adoption of the BitTorrent  P2P network \cite{BitTorrent}. Recently, the Interplanetary Files System (IPFS)\footnote{https://ipfs.io/}, a decentralized P2P distributed file system, has emerged with the objective to connect computers with the same file system and to distribute large datasets. IPFS can access files in any network by the file addresses, each of which is stored as a byte string. To better facilitate IPFS with credit incentives, filecoin\footnote{https://filecoin.io/}  is a token protocol whose blockchain runs on a novel consensus model, called Proof-of-Spacetime, where blocks are created by miners that store the data. The filecoin protocol provides a data storage and retrieval service via a network of independent storage providers that do not rely on a single coordinator, such that: 1) clients pay to store and retrieve data, 2) storage miners earn tokens by offering storage, 3) retrieval miners earn tokens by serving data. The filecoins can be exchanged for US dollars, Bitcoins, Ethereum, and more. In short, filecoin creates a decentralized storage network (DSN) and a cryptocurrency marketplace on top of it. 

\subsubsection{Data Sharing Credits} 

Similar sharing concept has been introduced into data/bandwidth sharing scenarios. RightMesh \cite{rightmesh} claims to be the world's first software-based, ad-hoc mobile mesh network that brings connectivity to all. The connectivity is in P2P mode via Wi-Fi, Bluetooth, and Wi-Fi Direct. When a client and hotspot node find each other, they form a new mesh for people to join and share, and it grows from there. Redundancy can strengthen the mesh network. In a densely-populated region, more available people and nodes can join the mesh network, which strengthens the robustness of the network. To encourage participation, a mesh node provider is awarded RMESH Tokens and the payment process is decentralized by leveraging the Ethereum platform \cite{rightmesh2}.

\subsubsection{Computational Sharing Credits}

At present, there is a growing need for computational power for scientific research, machine learning and graphics rendering in large ecosystems. This area has evolved from projects like BOINC \cite{boinc}, which relied on the goodwill of users to solve problems like DNA folding with their spare CPU cycles \cite{folding}. Some algorithms, such as machine learning and deep learning algorithms, and other sophisticated solutions are raising demands for high-performance hardware and more bandwidth to address the needs of enterprises and businesses in minutes \cite{computering_power}. To solve this problem, the idea of building a platform that enables participants to lend and borrow computing powers emerged. Golem\footnote{https://golem.network/} is a P2P platform that allows the participants to rent and buy computing powers directly by using cryptocurrency. In Golem, a distributed network of computers that are managed by blockchain and smart contracts is used to create an ecosystem where the computing power can be borrowed. Hong {\it et al.} in \cite{connectivity_d2d} proposed a connectivity-aware mobile computational resource sharing system in D2D networks. By incorporating a blockchain-empowered credit system, user selfishness in this D2D computational sharing system is effectively and significantly reduced  \cite{blockchain_d2d}.

\section{Desirable Characteristics of dApps} \label{sec:requirement}

According to the application scenarios discussed above, future dApps demand a blockchain platform that fulfills the following desirable characteristics:

\subsection{Better Performance}

\subsubsection{Low Latency}

Long transaction delay has been a critical issue since the birth of Bitcoin. Since the average time for the Bitcoin nodes to mine a block is 10 minutes, the average transaction confirmation time is around an hour (as a user typically waits for 6 blocks). Even though the response latency has been significantly reduced to around 15 seconds in Ethereum, a sufficiently small latency to support interactions of general applications is yet to be achieved. In fact, longer delays frustrate users and make dApps less competitive with existing non-blockchain alternatives. For instance, a common user in a blockchain-based social network website will typically require the system to respond to his/her like or share action to a post within 2 to 3 seconds.

\subsubsection{High Throughput}

Modern web-based systems, e.g., social networks, massive multi-player online games, online shopping malls, require the blockchain platform to support millions of active users on a daily basis. Therefore, the capability of handling a large amount of concurrent traffic is critical in a dApp platform. However, current blockchain platforms still suffer from throughput bottlenecks. For example, CryptoKitties, which gained a lot of popularity on its launch, at one point account for nearly 30\% of all transactions on Ethereum, which resulted in a peak backlog of about 30,000 pending transactions.

\subsubsection{Fast Sequential Performance} \label{sec:sequential}

In system designs, dependencies among software components or logical steps restrict the execution of an application. Some procedures in certain applications, such as updates on one particular piece of data, cannot be implemented in parallel, due to the sequential dependent on the results produced by previous steps. In blockchain systems, the sequential performance of a dApp is determined by the response delays from all nodes in the network, since all transactions/operations should be executed and verified by all nodes to reach a consensus. Therefore, the blockchain platform that hosts dApps needs fast sequential performance to handle high volumes. 

\subsection{Enabling Offline Transactions}

Many current blockchain systems depend on Internet connectivity in order to verify funds quickly. Systems participating in a particular blockchain network may go offline periodically. However, if a subset of devices disconnect from the Internet and exchange signed transactions with each other, there is no guarantee that double spending has not occurred if another device remaining online with the same key-pair as an offline device has the ability to simultaneously spend. For example, consider a group of people take a bus trip to a remote village with their mobile phones. The village has no Internet access. A dApp could be designed such that it could accept offline transactions which are signed for payment for goods. A person on the bus could send their payment for a coconut this way to a vendor using a local Bluetooth connection. When this signature eventually is relayed to the Internet at a later time, the payment would be successful, unless the person on the bus also had the same key-pair being used back in their home computer, and spent the money before they went offline. This problem becomes more complicated when large groups of devices fragment the network. Since many of the blockchains rely on over 51\% of devices to co-operate, there are potential malicious attacks possible whereby an attacker could attack the Internet infrastructure strategically in order to divide and conquer with 51\% attacks \cite{bitcoincentral}\cite{blockchainanomoly}.

\subsection{Reasonable Monetary Cost}

\subsubsection{Low Transaction Fee}

As part of the incentives for block producers, the concept of transaction fee was born with Bitcoin. In classic blockchain systems, e.g., Ethereum, transaction fees can also be a way to prevent spams or malicious executions of smart contracts, since intruders need to spend their tokens to start their attacks. However, transaction fees become a barrier for transactions with relatively small monetary values, due to the large proportion of the transaction overheads. In the current blockchain ecosystem, the dApp developers are struggling with the high transaction fees they need to pay when they deploy and execute their smart contracts. 

\subsubsection{Modern Free Internet Business Model}

Another critical issue related to transaction fees is the business model. By default, the action initiator, e.g., the invoker of the smart contract in Ethereum, need to purchase tokens before they can utilize the system. This limits the user base of the dApp, especially since cryptocurrency has yet to achieve universal acceptance in society. In fact, the modern Internet business model is based on the fast increase of user popularity, which implies that the dApp developers should have the flexibility to offer users free services. In other words, the users do not need to purchase or hold tokens to use the platform, which leads to more widespread adoption. Future dApp can adopt the modern Internet business model by offering free services to users and share the profit of the platform with its users and its content producers.

\subsection{Flexible Maintainability}

\subsubsection{Enabling System Upgrades}

As blockchain technologies are still in their infancy, it is inevitable that a blockchain system will require upgrades from one version to the next. However, due to the nature of P2P consensus, the hard fork is the only approach for current blockchain systems to upgrade themselves, which may result in the loss of participating network nodes. Another potential issue for a hard fork is that there will be multiple similar tokens sharing a common origin, which will confuse users. For example, like Bitcoin and Bitcoin Cash, 'Ethereum' (ETH) and 'Ethereum Classic' (ETC) forked from each other in July 2016. To this end, a system upgrade mechanism is needed for next-generation blockchain systems, which facilitate version control of dApps deployed over them.

\subsubsection{Easy Bug Recovery}

Security issues in smart contracts has been investigated in many previous works \cite{smartcontractdumb} \cite{SurveySmartContractAttack} \cite{SmartContractSecurityPattern}. Though most bugs and system flaws can be prevented by careful implementation and intensive tests, it is virtually impossible to guarantee that a non-trivial smart contract is bug-free. The situation is exacerbated by the high complexity of some dApps. However, the immutable nature of blockchain data prevents the modification of dApps, which makes the delivery of bug patch impossible. Therefore, the blockchain platform must provide flexibility in supporting bug recovery approaches for dApp developers, especially for those critical issues that may crush the whole ecosystem in dApps.

\subsection{Simpler Identity Management}

Many blockchain dApp systems are struggling with challenges around identity. Some systems such as ZCash\footnote{https://z.cash/} and Monero\footnote{https://getmonero.org/} try to hide the identity of users and transactions. There has also been recent work to add the ability for anonymity on top of existing blockchains, particularly in use-cases like Initial Coin Offerings (ICOs), where money is being fund-raised through smart contracts and regulatory bodies require the Know Your Customer (KYC) and the Anti Money Laundering (AML) checks \cite{anonblockchain} without giving up the identity of the contributors to the entire global network. On the other hand, there is a movement to create one common identity such as Blockstack\footnote{https://blockstack.org/} that can be used across all dApps in a similar way that openID\footnote{https://openid.net/} was used to create a common identity across web services.

\section{Considerations when selecting a Blockchain implementation}\label{sec:selection}

Different blockchain implementations with subtle differences in key technical areas are constantly emerging to address different shortcomings in existing systems. When selecting a potential blockchain technology, one may wish to have an implementation that is stable but may be willing to be flexible when necessary. This can be measured by looking at how often the network has "hard-forked" and how many derivative projects (forks on GitHub) exist. It is also desirable that the potential project has an active community of developers (internal and external) - which may be measured by metrics such as contributors, code commits and branches. Depending on the dApp, one may look for a blockchain technology that supports smart contracts and some form of scalable payments such as payment channels, as well as the economic model of the dApps being built on top, and has support for the correct programming languages for the project. To illustrate some of these considerations, the Bitcoin project and the Ethereum project are compared, however, any other project could be subjected to a similar comparison and analysis when selecting an appropriate technology for implementation.

\subsubsection{Bitcoin}

Bitcoin core's GitHub\footnote{https://github.com/bitcoin/bitcoin} lists 571 contributors and more than 18000 commits. There are many client implementations and APIs in a variety of languages with varying maturity. For instance, there is a Java library via the bitcoinj\footnote{https://github.com/bitcoinj/bitcoinj} project (and likely many others). The bitcoinj project has 95 contributors and more than 3000 commits. According to bitnodes\footnote{https://bitnodes.earn.com/}, there are about 10000 full nodes running Bitcoin (these are nodes running full verification of the entire blockchain transaction by transaction, as opposed to a thin client which relies on a full node which is trusted to do this on its behalf).  As of March of 2017, there were more than 10000 Bitcoin projects on GitHub\footnote{https://news.bitcoin.com/bitcoin-projects-github-surpass-10000/}. Bitcoin has been running since January 2009. According to GitHub, the Bitcoin source has been forked almost 20000 times, although the number of functioning forks that are active is likely much lower. The handling of actual forks as well as the market confusion and manipulations created after these forks make it difficult to select newly forked projects. According to blockchain info\footnote{https://www.blockchain.com/charts/n-transactions}, the highest 7-day average transactions per 24 hours seems to be about 425000, or 4.92 transactions per second (TPS). However, some studies have shown that it may be able to reach 7 TPS (with the 1MB block size) \cite{croman2016scaling}. The highest average transaction fee according to BitInfoCharts\footnote{https://bitinfocharts.com/comparison/bitcoin-transactionfees.html} was around USD \$55. With such a low transaction rate and high transaction fee, it is clearly not feasible to create transactions at a very granular rate for dApps (which severely restricts the type of applications that are possible without a scalable payment solution).

\subsubsection{Ethereum}
The Ethereum project has a few key GitHub repositories. As of August 2018, the go-ethereum repository\footnote{https://github.com/ethereum/go-ethereum} has 318 contributors, the cpp-ethereum repository\footnote{https://github.com/ethereum/cpp-ethereum} has 136 contributors, ethereum-j\footnote{https://github.com/ethereum/ethereumj} has 69 contributors (and 5012 commits). Solidity\footnote{https://github.com/ethereum/solidity}, which is one of the smart-contract languages in Ethereum, has 263 contributors. In total across these repositories, there are about 60000 commits. It's likely that some of the contributors overlap from the different parts of the project, but it is safe to say that Ethereum is at least comparable to Bitcoin in terms of the number of developers working on the project. According to ethernodes\footnote{https://www.ethernodes.org/network/1} there are 16000 full nodes running Ethereum. Due to the way Ethereum is organized into different projects, it is difficult to get one number for the number of forks (like contributors). For instance, go-ethereum has 6800 forks, cpp-ethereum has 2000, and ethereumj has 890. Ethereum has been active since July 30th, 2015. According to etherscan\footnote{https://etherscan.io/chart/tx}, the highest number of transactions per 24-hour period was 1,349,890 or 15.62 transactions/second (almost four times more than Bitcoin or twice as much as the 7 transactions/second Bitcoin should be able to reach). Rouhani and Deters showed that Ethereum transaction speed depends on which client implementation is used, with the parity client performing significantly better than the geth client \cite{ethscaling}. The highest average transaction fee according to BitInfoCharts\footnote{https://bitinfocharts.com/comparison/ethereum-transactionfees.html} was around USD \$4.15. Again there are similar concerns with respect to Bitcoin in regard to being able to execute transactions at fine granularity without overwhelming the network transactions throughput and paying more to settle transactions compared to the value of the data being sent.

\subsubsection{Other Blockchains}
In general, there are many forks of these two projects to choose from, and many other new takes on blockchains. Many of these projects have not yet undergone the scrutiny that the main chains like Ethereum and Bitcoin have undergone. There are very few analyses by independent parties that examine things like the theoretical limits of transactions throughput, in-depth security audits, economics and business models, and a multitude of other concerns. Many have underdeveloped communities which may not persist, with unclear roadmaps. Developers of dApps should consider both technical suitability as well as the long-term stability of the projects before choosing a particular technology for the development. Presently, this type of analysis must be done by the dApp developers, but there is incredible potential for the research community to critically evaluate the options available, to highlight best practices, what to avoid, how to improve, and how to achieve scalability and sustainability.

\section{Recent Developments in Blockchain Systems} \label{sec:development}

In order to support above desirable characteristics of dApps, both academia and industry have spent tremendous resources and efforts in developing next-generation blockchain systems. In this section, we will summarize state-of-art research directions in this area.

\subsection{Payment Channels and Payment Networks}
 Cryptocurrencies on blockchains work by recording every transaction on blockchains. It has many unique features like decentralization, transaction transparency and so on, but it has severe problems in terms of scalability. When there is a burst of transactions, it takes too long to write all backlogged transactions into a blockchain, especially for the blockchain built atop PoW. In PoW, creating every single block needs huge computing power. In order to reduce the number of transactions that hit the blockchain, the concept of payment channel is proposed. A payment channel is designed to facilitate the payment between two parties, which allows users to make multiple payments without triggering multiple transactions. In general, a payment channel can be either unidirectional or bidirectional. In this part, we will first introduce the unidirectional payment channel and then introduce the bidirectional one. After that, we will briefly introduce the payment network. 
 
 \subsubsection{Unidirectional Payment Channels}

\begin{figure}[htp]
\centering
\includegraphics[width=85mm]{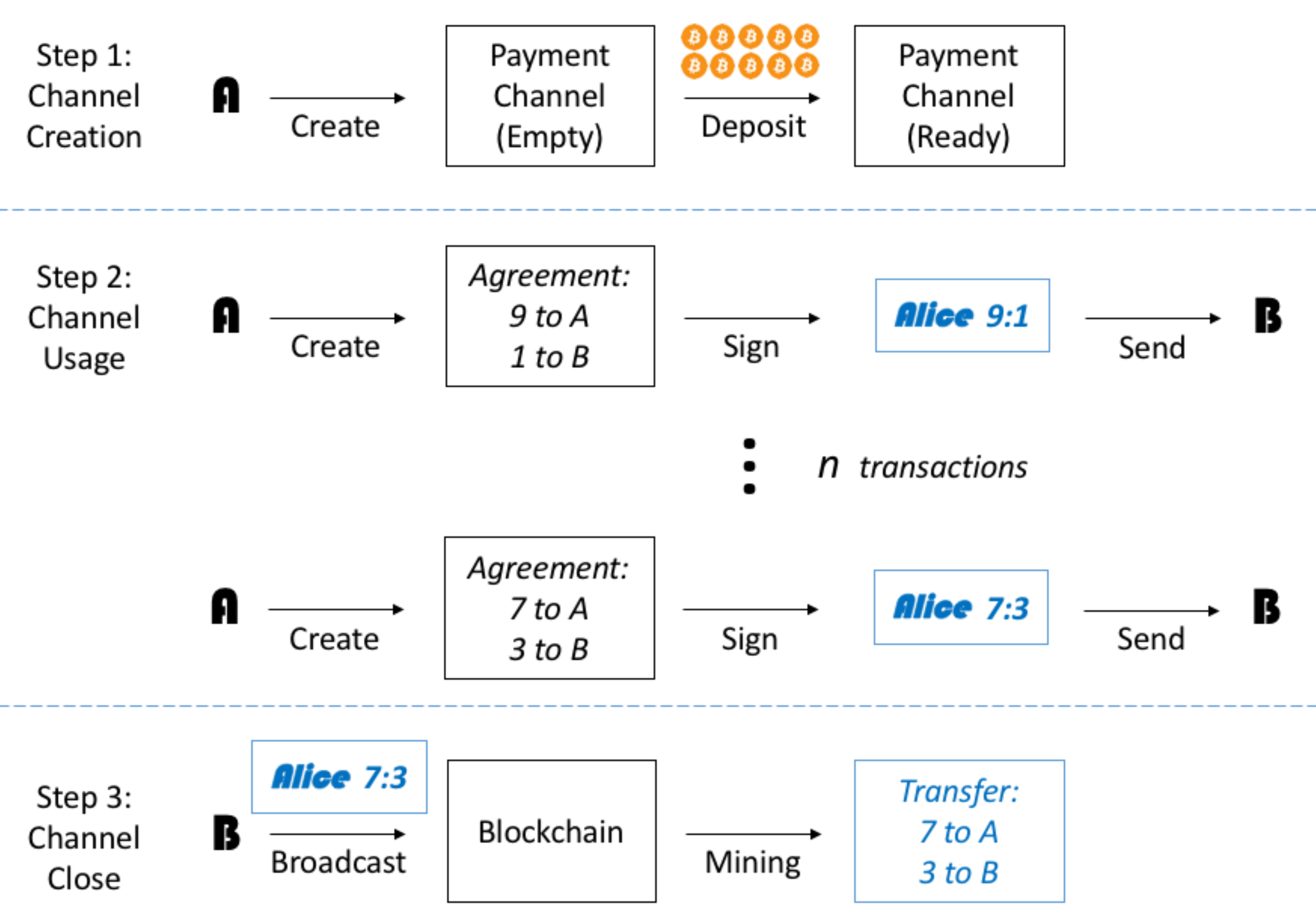}
\caption{The Life Cycle Unidirectional Payment Channel}
\label{fig:unidirectional}
\end{figure}

 For the ease of presentation, we assume that user $A$ needs to pay some cryptocurrency to user $B$ multiple times over a period of time. The trivial solution to handle multiple payments from $A$ to $B$ is that whenever $A$ wants to pay $B$, $A$ first signs a transaction with the payment amount and then broadcasts the signed transaction to the P2P network. The transaction will be recorded and confirmed. In other words, if user $A$ pays user $B$ say $n$ times in a period of time, $n$ transactions will be generated by $A$ and mined by miners. If users $A$ and $B$ do not interact with other users\footnote{This assumption is not necessary for practical payment channel implementation, which will be discussed later.}, people need only be informed about the final balances of $A$ and $B$ once. The payment channel is proposed by following the above logic. In particular, a payment channel is like a joint banking account where the cryptocurrency inside can be split and transferred into two wallets. In the context of unidirectional payment channel, since $A$ pays $B$ only, it is the responsibility of $A$ to create the payment channel and lock some deposit in it as shown by Step 1 in Figure\;\ref{fig:unidirectional}.

 Whenever $A$ would like to pay $B$, rather than creating a transaction and broadcasting it to the P2P network, $A$ can sign a signature splitting the cryptocurrency in the payment channel and send $B$ the signature, as shown in Step 2 of Figure\;\ref{fig:unidirectional}. Note that, when $B$ receives the signature, $B$ has not received the cryptocurrency yet. It is because the signature is not broadcast on the P2P network and the cryptocurrency in the ledger is not split yet. However, $B$ could get paid whenever he/she would like to by sending the signature to the blockchain. At this time, we say the payment channel is closed, as shown in Figure \ref{fig:unidirectional} at Step 3.

 The advantage of using payment channels is that, since $A$ may pay $B$ multiple times, $B$ can just wait for another signature from $A$. In the unidirectional payment channel, the latterly signed signature from $A$ is always more preferable by $B$, so $B$ can wait until $A$ sign the $n^{\text{th}}$ signature and broadcast the latest signature to the blockchain. Meanwhile, since the deposit is already transferred out from $A$'s wallet and locked into the payment channel, which will not be split until the channel is closed, $A$ and $B$ can still interact with other users and no conflict will occur. In summary, $A$ and $B$ have just two transactions on the blockchain: the transaction of $A$ creating the ledger and putting a deposit in it, and the transaction of $B$ broadcasting the final signature. However, $A$ can actually pay $B$ as many times as he/she wants to. We want to highlight that the easiest way to implement the unidirectional payment channel is only allowing $B$ to close the payment channel since $B$ is the unidirectional receiver who does not have the incentive to lie. When $B$ closes the payment channel by broadcasting the latest signature from $A$, the remaining amount of cryptocurrency in the payment channel not used by $A$ will be reimbursed to $A$'s wallet. 
 
 To also enable user $A$, a.k.a. the payer, to close the payment channel, we need to associate every signature of $A$ with a time-stamp and introduce the ``challenging period'' to the mechanism. When $A$ broadcasts his/her own signature into the blockchain, the time-stamp of the signature will be logged and the payment will not be closed immediately but going into the challenging period. If user $B$ had received $A$'s signature with a newer time-stamp, he can broadcast the newer signature and the previous one will be overwritten. The overwriting process can repeat between $A$ and $B$ until the end of the challenging period. Eventually, the cryptocurrency in the payment channel will be split with the last signature broadcast in the P2P network.
 
 \subsubsection{Bidirectional Payment Channels}
 After the unidirectional payment channel has been introduced, it will not be difficult to understand the bidirectional payment channel. As the name indicates, if there is a bidirectional payment channel between users $A$ and $B$, each of them can pay the other by signing a signature and sending the signature to the other party. The prerequisite of using a bidirectional payment channel between users $A$ and $B$ is that both of them need to contribute to the deposit. For example, let us assume $A$ and $B$ have contributed $5$ dollars each in the payment channel, so there are $10$ dollars in the bidirectional payment channel. When $A$ wants to pay $B$ $2$ dollars, $A$ needs to sign a signature of splitting the $10$ dollars. In our case, the splitting plan signed by $A$ is $7$ dollars going to $B$'s wallet and $A$ having the remaining $3$ dollars. When $B$ needs to pay $A$ $1$ dollar, $B$ can create another signature that $4$ and $6$ dollars are going to $A$ and $B$'s wallets, respectively. Furthermore, the deposit that a user contributed to the payment channel is the maximum payment that he/she can pay its opponent. Moreover, it is an intrinsic requirement that either user of the payment channel can close the channel when he/she wants to. However, people may not tell the truth. Reviewing the example given above carefully we can find that user $B$ can still close the bidirectional payment channel by broadcasting the signature signed by $A$ in the first round, so it seems like $B$ could get $7$ instead of $6$ dollars if he/she is a liar! 
 
 In order to deal with this situation, let us recall what we have done with the unidirectional payment channel when we allow both users to close the payment channel. We introduced the time-stamp and challenging period. So, if one of the users, say $A$, finds that user $B$ tried to close the payment channel dishonestly (as the payment channel has entered the challenging period and the existing split plan reported by $B$ is unfair to $A$), then $A$ can broadcast the signature signed by $B$. As the $B$'s signature has a later time-stamp than $A$'s signature, the splitting plan of cryptocurrency in payment channel will be updated. Similar to the unidirectional payment channel when allowing both users to close the channel, the splitting plan is locked after the challenging period and either of them can finally close the payment channel and the cryptocurrency flows to each one's wallet.
 
 \subsubsection{Payment Networks}
To better understand payment networks, we can draw an analogy between payment networks and communication networks. The link layer in a communication network is very similar to the payment channel in a payment network, while the end-to-end communication in a communication network is just like the multiple hop payment in a payment network. The reason why we want to use a payment network is that creating a payment channel, no matter whether it is unidirectional or bidirectional, still requires updating on the blockchain. If there is another user that has set up payment channels to other users, this user can relay the payment. For example, if there are two payment channels such that one is between users $A$ and $B$ and the other is between users $B$ and $C$. When $A$ wants to pay $C$, say $2$ dollars, he can simply pay user $B$ and let $B$ pay user $C$. The problem is that $B$ or $C$ can lie. For example, after $B$ receives the payment from $A$, $B$ may refuse to pay $C$. Or, $C$ may say that he did not receive any payment from $B$ even though he did.

The basic principle to solve the problem is letting $A$ first create a puzzle and send the key of the puzzle to $C$. The puzzle is very difficult to solve, but it is very easy to validate the key, like the {\it hash} operation. Then, $A$ gives $B$ the puzzle and reach an agreement with $B$ as follows: ``if $C$ offers you ({\it i.e.} user $B$) the correct key, send $C$ $2$ dollars and I ({\it i.e.}  user $A$) will reimburse you ({\it i.e.} user $B$) $2$ dollars when you ({\it i.e.} user $B$) tell me what the correct answer is.''  
 
\subsubsection{Limitations of Payment Channels}
 
Would payment channels be the ultimate solution? The answer is not for all dApp scenarios. As discussed in Section \ref{sec:sequential}, sequential dependencies on the data resulting from previous steps are essential requirements for many software implementations. The off-chain data caching nature of payment channels will prevent the data from being synchronized by all components of the system. Therefore, payment channels are not yet perfect in supporting next-generation dApps.
 
\begin{table*} [htbp]
  \renewcommand{\arraystretch}{1.1}
  \caption{Comparison among Different Consensus Models}
  \begin {center}
  \begin {tabular}{| p{0.8in}|  p{1.7in} | p{1.7in} | p{2.1in} | }
  \hline 
  \hline \centering
   & {\bf PoW} & {\bf PoS} & {\bf DPoS} \\
  \hline \centering
  Metaphor & City State Democratic System & Capitalism System & Parliamentary System \\
  \hline \centering
  Mechanism & One CPU One Vote & One Token One Vote & Vote for Delegates \\
  \hline \centering
  Block Rewards & To Miners Solved PoW & To Token Holders as Interest & To Elected Supernode Producing Blocks \\
  \hline \hline
  \end{tabular} \label{tab:comparition}
  \end{center}
\end{table*}

\subsection{Novel Consensus Models}

Though the creative application of PoW consensus model started the new era for blockchain, it is also criticized for its energy inefficiency nature: all participating nodes in the PoW network are doing useless mathematical work for the privilege of writing blocks, which costs a tremendous amount of electricity.  For example, the annual energy consumption index for Bitcoin mining alone is 11.8\% more than that of Switzerland, and $\sim$30\% that of Australia with a landmass of more than 7 million square kilometres\footnote{https://digiconomist.net/bitcoin-energy-consumption}. Also, note that this energy consumption is still growing fast for Bitcoin at a rate of $>$500\% from May 2017 to May 2018. In fact, recent research \cite{bitcoin_pow} predicts that Bitcoin transactions may consume as much electricity as Denmark by 2020. Moreover, adopting PoW is also the intrinsic reason for high transaction fee and long latency. Therefore, investigating an efficient consensus model for future blockchain systems has been a hot topic in both academia and industry. In this section, we review some recent novel consensus models.

\subsubsection{Proof of Stake (PoS)}

As we revealed in Section \ref{sec:pow}, PoW leverages hardware investment to prevent identity forges in Sybil attacks. In contrast, the PoS consensus model\footnote{https://bitcoinmagazine.com/articles/what-proof-of-stake-is-and-why-it-matters-1377531463/} tries to find an alternative solution to this problem. Different from PoW, the network participants need not solve mathematical problems in order to write a block. Instead, the producer of a block is randomly chosen based on the participant's ownership of stake (i.e., the more stake a participant has, the more likely it can become a block producer). Under this circumstance, the amount of tokens one node holds becomes the barrier of the identity forge. In other words, the system intruders will need to hold a majority of the coins in circulation to perform 51\% attack. In fact, this is extremely difficult: due to the laws of supply and demand, the price of tokens in a system will continuously increase when the intruders start their purchase, which may punish them economically. More interestingly, once the intruders become the major stakeholders of a digital currency, they lose their motivation to attack: their attack will disrupt the operation of the currency, which in turn introduces financial damage to the intruders. From another perspective, the PoS is similar to PoW in terms of creating block producing barrier. The only difference is that PoS encourages network participants to invest their money on tokens, rather than mining machines. So does PoS solve the tremendous overhead introduced by mathematical problem-solving in PoW while preventing Sybil attacks? The answer is affirmative. However, it does not mean that PoS is the perfect consensus model. One critical issue in PoS is the rational forks by the stakeholders. As we discussed, PoS utilizes stake to replace the PoW computation. However, once a block producer in a PoS blockchain creates a fork, there is no cost for all stakeholders to follow the sub-chain spontaneously. Technically, one fork will double the stakeholders' tokens and two forks will triple them. There is nothing to lose for the stakeholders to follow all chains and receive multiple coins in different sub chains. Too many forks on one blockchain will introduce chaos and confusions, thereby reducing the value of the network. Due to these considerations, only a few cryptocurrencies available in the market are based on PoS, such as Peercoin\footnote{https://peercoin.net/} and ShadowCash\footnote{https://github.com/shadowproject/shadow}.

\subsubsection{Delegated Proof of Stake (DPoS)}

The DPoS consensus model,  as explained in ``DPOS Consensus Algorithm - The Missing White Paper'' for STEEM\footnote{https://steemit.com/dpos/@dantheman/dpos-consensus-algorithm-this-missing-white-paper}, solves the identity forge problem from another aspect: network participants delegate their rights of producing blocks to a small group of supernodes. The way that DPoS creates barriers for identity forge in Sybil attack is the difficulty of becoming a supernode. In a typical DPoS consensus, the stakeholders need to vote for their preferred block producer candidates, and those successfully elected receive rewards from creating correct and timely blocks. With DPoS, the computational overhead for PoW is eliminated since the block producers do not have to compete with each other in mathematical computations. Also, the stakeholders cannot perform rational forks, since the votes allocated to the stakeholders are limited in quantity, e.g. proportional to the number of tokens they hold. On the other hand, the elected block producers are supervised by the majority of stakeholders to perform their duties for the incentives generated by creating new blocks. Any malicious behaviors from block producers will be reported and unqualified block producers will be voted out as a consequence. The number of block producers is subject to different implementations. For example, EOS\footnote{https://www.eos.io/} has 21 supernodes while Asch\footnote{http://www.asch.so/} has 101 delegates. Block producers may also serve as governance gateway. Any proposed change on system parameters, such as transaction fee, block size, witness pay or block intervals, needs to be approved by a majority of block producers. Since there is only a limited number of block producers in DPoS, and the voting procedure can readily screen out low-quality candidates, it is easier for the system to optimize itself in terms of performance. Accordingly, DPoS features relatively low latency, high efficiency, and flexibility. However, there are doubts around the mechanism of delegated block producers: opponents criticize that DPoS is not a decentralized platform since it is impossible to guarantee the purity of block producers. The small group of block producers may conspire to maximize their own interests. Also, since the block producers will receive rewards, a group of candidates who did not get elected may create forks on the main chain, which results in multiple chains as well.
In summary, DPoS proposes to leverage the power of stakeholder approval voting to resolve consensus issues in a fair and democratic way.

\subsubsection{Comparison Among Consensus Models}

Table \ref{tab:comparition} provides a comparison among different consensus models. We would like to utilize three political models as the metaphor for PoW, PoS, and DPoS. As the first generation blockchain system, PoW is the original P2P consensus model for blockchains, which is analogous to democratic voting in ancient European city-states. Its ``One CPU One Vote'' idea is exactly the same to the ``One Man One Vote'' form. However, once the size of the system increases to a certain level, this form of democracy becomes inefficient. On the other hand, PoS borrows the idea of interest produced by cash savings, so that newly generated tokens are distributed to those stakeholders in proportion to their current holdings. More tokens indicate more benefits in the system, which is a feature of the capitalist systems: the means of production derives a passive income from their operation. In contrary, DPoS borrows from the political model of parliamentary systems adopted by many countries: representatives are elected by the public to efficiently solve the legal and social issues. Most blockchain systems allow a certain amount of inflation for the circulating tokens. A common practice is to generate new coins as block rewards for block producers, which encourages the participants of the system. Due to their unique properties, different consensus mechanisms should be associated with different reward models, as listed in Table~\ref{tab:comparition}.

There is still significant ongoing research on creating novel consensus models. Recent proposals include Leased Proof of Stake\footnote{http://wiki.p2pfoundation.net/Leased\_Proof\_of\_Stake}, Proof of Burn\footnote{http://slimco.in/}, Proof of Capacity\footnote{https://www.burst-coin.org/}, Proof of Elapsed Time\footnote{https://nulltx.com/what-is-proof-of-elapsed-time/}, Algorand\footnote{https://www.algorand.com/}, etc. However, these protocols have yet to achieve wide acceptance by the dApps community.

\subsection{Beyond Public Blockchains}
Public blockchains are also referred to as permissionless blockchains as system participants do not need any permission before joining the network. In some application scenarios where transaction frequency or data privacy is critical, e.g., certain decentralized high update rate enterprise record-keeping applications or storage of medical records, permissionless blockchains are challenged by their low efficiency and highly open nature. 

First, most permissionless blockchains have significant bottlenecks on efficiency (typically in terms of TPS) where the necessary level of security is based on having a large number of network participants, such that network synchronization (or consensus) alone already limits the TPS. Moreover, most of the permissionless blockchains online are PoW-based. Therefore, even if a certain level of TPS requirement is met, it comes at a price of huge consumption and waste of energy.

Openness can yet be another issue of permissionless blockchains for a decentralized medical recording application. Even though privacy is to some extent provided by permissionless blockchains in the way of anonymizing transacting parties, many transactions can still be linked, potentially resulting in speculation and/or manipulation of users' privacy. For example, a user of this type of decentralized system, e.g., when applied to medical record keeping, may be identified by her colleagues by comparing the time she is away from work and timestamps of recent transactions. It is even worse if malicious parties find a security hole in the smart contracts of the medical record keeping application, which may result in horrible privacy breaches. 

Unlike the permissionless blockchains, permissioned blockchains have restrictions on network participation. Specifically, permissionless blockchains like Bitcoin and Ethereum allow anyone to read records on blockchains, to make transactions, or to become a miner, while specific invitations are needed to participate in a permissioned blockchain, e.g., HyperLedger Fabric\footnote{https://www.hyperledger.org/projects/fabric}. 

Many users find it difficult to differentiate permissioned blockchains from permissionless ones due to their similarities:
\begin{itemize}
    \item Both are decentralized and P2P;
    \item Both participants share the same copy of append-only ledger of transactions;
    \item Both participants synchronize the network through consensus;
    \item Both try to provide a certain level of the immutability of the shared ledger, etc.
\end{itemize}
Further confusion is caused by the evolution of permissioned blockchains over the past years. In general, permissioned blockchains can be categorized into two broad types: private blockchains and consortium blockchains. Private blockchains have the strictest system participation control. All reading, transacting, and mining privileges are strictly controlled within a single organization by the network owner. In comparison, consortium blockchains are subtly different from private blockchains in that the system participation is controlled by a number of organizations that form the consortium. 

In the area where permissionless and permissioned blockchains overlap there are hybrid blockchains. Hybrid blockchains try to combine the advantages of both permissionless and permissioned blockchains, compromising among security, efficiency, cost, fairness, etc., to meet the increasingly complex application requirements.

In this section, we explain these different types of blockchains through examples.

\begin{table*} [htbp]
  \renewcommand{\arraystretch}{1.1}
  \caption{Comparison among Different Blockchains}
  \begin {center}
  \begin {tabular}
  {| p{1.2in} | p{1.2in} | p{1.2in} | p{1.2in} | p{1.2in} |}
  \hline
  \hline
  {\bf Parameters} &
  {\bf Permissionless} & {\bf Private} & 
  {\bf Consortium} & {\bf Hybrid} \\
  
  \hline
  \bf Network &
  Decentralized & Centralized & 
  Centralized & Hybrid \\
  
  \hline
  \bf TPS &
  Low & High &
  High & High \\
  
  \hline
  \bf Visibility \& Participation &
  Open & Restricted &
  Restricted & Varies \\
  
  \hline
  \bf System Governance &
  Hard & Easy &
  Medium & Varies \\
  
  \hline
  \bf Security &
  Varies & High &
  High & High \\
  
  \hline
  \bf Examples &
  Bitcoin, Ethereum, EOS & MultiChain &
  Hyberledger Fabric & XinFin \\
  
  \hline
  \hline
  \end{tabular} \label{tab:chain comparison}
  \end{center}
\end{table*}

\subsubsection{Private Blockchains}
A private blockchain has access control and operates under a specific organization. Participants need to be invited, and existing participants may decide on future entrants. Once an entity has joined the network, it will play a role in maintaining the blockchain in a decentralized manner. In addition, private blockchains rely on internal participants' honesty to verify transactions, which saves the efforts and potential wastage of mathematical PoW as the means of maintaining security. Overall, private blockchains are more efficient in terms of scalability and compliance.

MultiChain\footnote{https://www.multichain.com/}, as an example of private blockchains, is a platform that helps organizations to build a private blockchain for financial transactions. In traditional blockchains, access to a private key means the ownership of the funds. 
In contrast, beyond using only private keys to control the funds, MultiChain has developed the ``handshaking" process in its whitepaper \cite{MultiChainWhitePaper}.
The process needs two participants to first connect, and then verify permission of each other to enter inter a transaction. 
After verification, they send each other a challenge message, which is returned with a signature to prove the ownership of the funds. If an agreement is not reached, the connection will be aborted.

Furthermore, MultiChain has resolved a notorious dilemma posed by most private blockchains, i.e., a participant may monopolize the mining process.
The solution lies in the introduction of a parameter called "\emph{mining diversity}", restricting the number of blocks that may be produced by the same miner within a given time window. If the miner of a new block is proven to have violated the requirement of \emph{mining diversity}, this block will be deemed invalid by the network. Consequently, the higher the \emph{mining diversity} is, the less chance that a miner could monopolize the network.

Overall, MultiChain has the following desirable characteristics:   
\begin{itemize}
 \item enabling secure mining without expensive PoW consensus that leads to enormous power waste, which meanwhile enhances scalability;
 \item enabling network administrators to manage privileges of upcoming participants;
 \item preventing the network from being monopolized by a miner by introducing \emph{mining diversity} such that a miner cannot over-produce too many blocks within a time window.
\end{itemize}

\subsubsection{Consortium Blockchains}
From some people's perspectives, consortium blockchains are a subset of private blockchains. Therefore, they are also called "partially private".Similarly, it features many of the same benefits as private blockchains, such as high efficiency, high scalability, and greater transaction privacy than permissionless blockchains. However, rather than having an organization in full control of the blocks, the consortium blockchain is a blockchain where the consensus process is controlled by a pre-selected set of nodes; e.g., at least 10 out of 15 organizations in this consortium need to sign and approve a block for it to be valid. It solves the problem of private blockchains that they are more vulnerable to being hacked and information altering in internal networks.

Hyperledger Fabric is an example of consortium blockchain implementation for distributed ledger solutions. In Hyperledger Fabric, the consensus consists of 3 phases implemented by participating nodes from different organizations:
\begin{enumerate}
    \item Endorsement: to get at least $m$ out of $n$ participants' signatures to endorse a transaction.
    \item Ordering: accept the endorsed transactions and agree to the order to be committed to the ledger.
    \item Validation: validate the results of ordered transactions and check endorsement policy and double-spending.
\end{enumerate}
This has implemented a better division of labor, and the applications may choose different endorsement, ordering, and validation based on their different needs. In addition, Fabric has fewer nodes than permissionless blockchains and computes data massively in parallel, which makes Fabric's scalability much greater than the permissionless blockchains. Indeed, Fabric can scale to over 1000 TPS in a very short time. Overall, Fabric as an example of consortium blockchain strengthens its flexibility in security and permission.

\subsubsection{Hybrid Blockchains}
As we discussed above, the consensus of a permissioned blockchain is controlled by one or several parties, and consensus of a permissionless blockchain is not controlled by any party but agreed by a majority of the users in the network. Hybrid blockchains are the combination of these two types. 
It can make the transactions private but still verifiable by an immutable record on the permissionless blocks.

An example of a hybrid blockchain is XinFin\footnote{https://www.xinfin.org/}, which aims to bridge the \$5 trillion global infrastructure deficit by letting institutions and/or governments connect blockchain-based digital assets to IoT enabled equipment in order to raise foreign direct investments and enable peer-to-peer financing. XinFin foundation is a non-profit organization which liaises with different international governments in order to reduce the existing gap in global infrastructure. According to XinFin, the lack of government-sponsored financing hinders the possibility of many infrastructure projects around the globe. However, by creating a secured blockchain transaction platform, XinFin aims to bridge that gap wherein investors can bid for different infrastructure projects and finance them in a smoother way, thereby avoiding all the issues and paperwork that arise from finance an infrastructure project across different countries.

To sum up, Table~\ref{tab:chain comparison} depicts the different traits, favorable application scenarios, and examples for the different flavors of blockchains.

\section{Conclusion}\label{sec:conclusion}

Blockchain systems leverage cryptography technologies, P2P networking and consensus models to provide infrastructures for decentralized applications. In this article, we have reviewed the history of blockchain systems and clarified their common definitions. We have presented the application scenarios of dApps, which in our opinion is the subject matter of future blockchains. We have also discussed the desirable characteristics of dApps and recent directions in blockchain development, including payment channels, novel consensus models and non-public blockchains. We believe that networked computing systems are on the edge of a new era of the decentralized ecosystem, which will eventually lead to the next-generation Internet services.

\bibliographystyle{ieeetr}
\bibliography{library}

\begin{IEEEbiography}[{\includegraphics[width=1in,height=1.25in,clip,keepaspectratio]{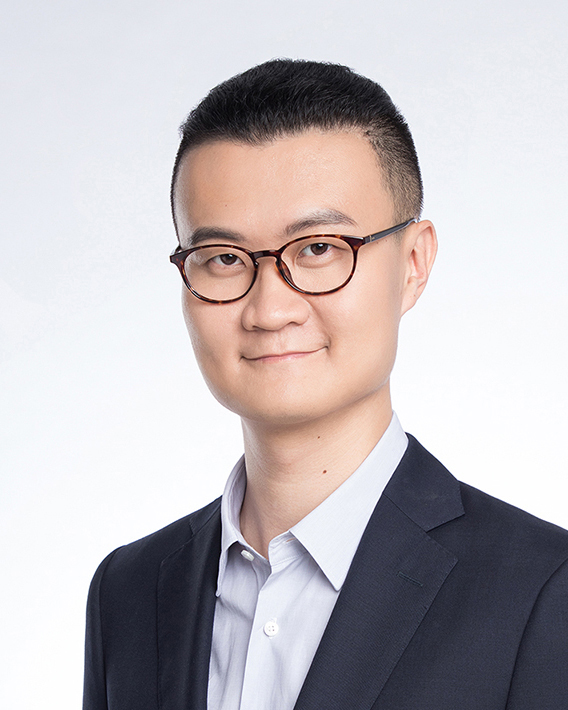}}]{Wei Cai} [S'12-M'16] received the B.Eng. degree in Software Engineering from Xiamen University, China in 2008, the M.S. degree in Electrical Engineering and Computer science from Seoul National University, Korea, in 2011, and the Ph.D. degree in Electrical and Computer Engineering from The University of British Columbia (UBC), Vancouver, Canada, in 2016. 
From 2016 to 2018, he was a Postdoctoral Research Fellow with UBC. He joined the School of Science and Engineering, The Chinese University of Hong Kong, Shenzhen, in August 2018, where he is currently an Assistant Professor. He has completed visiting research at National Institute of Informatics, Japan, The Hong Kong Polytechnic University, and Academia Sinica, Taiwan. His recent research interests include software systems, cloud and edge computing, blockchain systems, and networked video games. 
Dr. Cai was a recipient of the 2015 Chinese Government Award for the Outstanding Self-Financed Students Abroad, UBC Doctoral Four-Year-Fellowship from 2011 to 2015, and the Brain Korea 21 Scholarship. He also received the best paper awards from CloudCom2014, SmartComp2014, and CloudComp2013.
\end{IEEEbiography}

\begin{IEEEbiography}[{\includegraphics[width=1in,height=1.25in,clip,keepaspectratio]{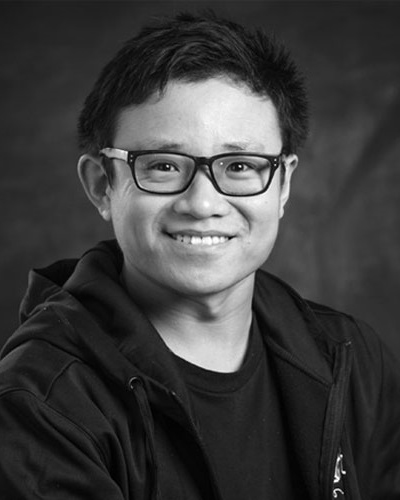}}]{Zehua Wang} [S'11-M'17] received the B.Eng. degree in Software Engineering from Wuhan University, Wuhan, China, in 2009, the M.Eng. degree in Electrical and Computer Engineering from Memorial University of Newfoundland, St. John's, NL, Canada, in 2011, and the Ph.D. degree from the University of British Columbia (UBC), Vancouver, BC, Canada, in 2016. He is currently a postdoctoral research fellow at UBC, Vancouver, BC, Canada, and the Chief Micropayments Scientist in RightMesh Project, BC, Canada. His research interests include blockchain technology, system optimization, social networks, and mobile ad hoc networks. He received the Chinese Government Award for Outstanding Self-Financed Students Abroad in 2015. He was the recipient of the Four Year Doctoral Fellowship (FYF) at UBC from 2012 to 2016. He was also awarded the Graduate Support Initiative (GSI) Award at UBC in 2014 and 2015. Dr. Wang served as the technical program committee (TPC) Co-chair of IEEE International Workshop on Smart Multimedia (SmartMM '17, '18).
\end{IEEEbiography}

\begin{IEEEbiography}[{\includegraphics[width=1in,height=1.25in,clip,keepaspectratio]{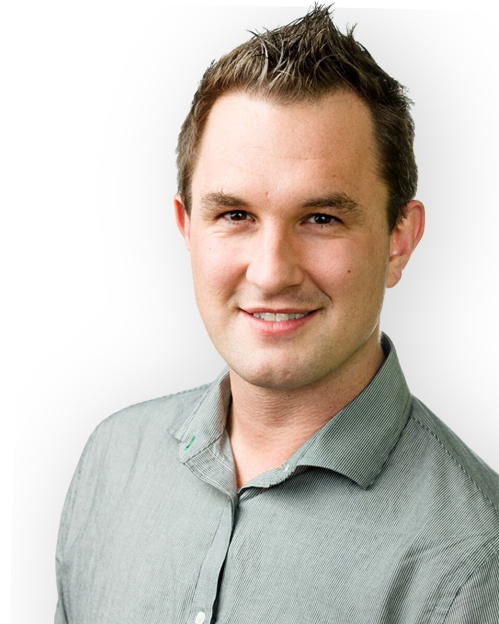}}]{Jason B. Ernst} [S'09-M'16] is the CTO of RightMesh and holds a PhD and M.Sc both in Applied Computing at the University of Guelph in Guelph, Canada (2016, and 2009 respectively). Jason is also an adjunct professor at the University of Guelph since 2017. Jason obtained an Honours B.Sc. in Computer Science from Wilfrid Laurier University, in Waterloo, Canada in 2007. Jason has more than 30 peer-reviewed, published papers on wireless networks, cognitive agents, FPGAs, and soft-computing topics and has presented his research at international conferences around the world. Prior to joining RightMesh, Jason was the CTO of Redtree Robotics where he focused on mesh networks to enable swarm robotics. Jason is an inaugural member of the ACM Future of Computing academy and has served as TPC and technical reviewer on many international journals, conferences, and workshops.
\end{IEEEbiography}

\begin{IEEEbiography}[{\includegraphics[width=1in,height=1.25in,clip,keepaspectratio]{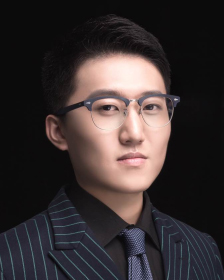}}]{Zhen Hong} [S'15] received his B.A.Sc. in Electrical and Computer Engineering from The University of British Columbia (UBC), Vancouver, Canada in 2015. He is currently an M.A.Sc. student in UBC and a research assistant in the Wireless Networks and Mobile Systems (WiNMoS) Laboratory led by Prof. Victor C.M. Leung at UBC. His recent research interests include blockchain technology, mobile cloud computing, system, and network design, and modeling. He received the best paper award at SmartComp2014 held in Hong Kong. He is a student member of IEEE.
\end{IEEEbiography}

\begin{IEEEbiography}[{\includegraphics[width=1in,height=1.25in,clip,keepaspectratio]{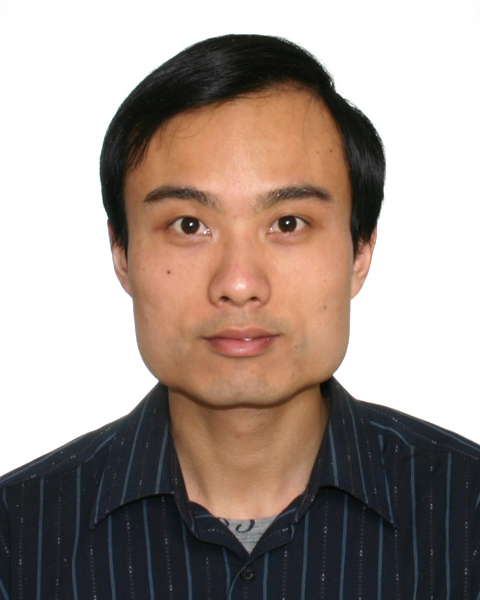}}]{Chen Feng} [M'15] received the B.Eng. degree from the Department of Electronic and Communications Engineering, Shanghai Jiao Tong University, China, in 2006, and the M.A.Sc. and Ph.D. degrees from the Department of Electrical and Computer Engineering, University of Toronto, Canada, in 2009 and 2014, respectively.
From 2014 to 2015, he was a Postdoctoral Fellow with Boston University, USA, and Ecole Polytechnique Federale de Lausanne (EPFL), Switzerland. He joined the School of Engineering, University of British Columbia, Kelowna, Canada, in July 2015, where he is currently an Assistant Professor. His research interests are in coding theory and its applications in various fields, ranging from wireless communications to quantum communications, and from communication networks to blockchain systems.
Dr. Feng was a recipient of the prestigious NSERC Postdoctoral Fellowship in 2014. He was recognized by the IEEE Transactions on Communications (TCOM) as an Exemplary Reviewer in 2015. He is a member of ACM and IEEE.

\end{IEEEbiography}

\begin{IEEEbiography}[{\includegraphics[width=1in,height=1.25in,clip,keepaspectratio]{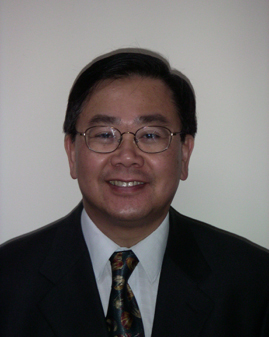}}]{Victor C.M. Leung}
[S'75-M'89-SM'97-F'03] is a Professor of Electrical and Computer Engineering and holder of the TELUS Mobility Research Chair at the University of British Columbia (UBC).  His research is in the broad areas of wireless networks and mobile systems. He has co-authored more than 1200 journal/conference papers and book chapters. Dr. Leung is serving on the editorial boards of the IEEE Transactions on Green Communications and Networking, IEEE Transactions on Cloud Computing, IEEE Access, IEEE Network, and several other journals. He received the IEEE Vancouver Section Centennial Award, 2011 UBC Killam Research Prize, 2017 Canadian Award for Telecommunications Research, and 2018 IEEE TGCC Distinguished Technical Achievement Recognition Award. He co-authored papers that won the 2017 IEEE ComSoc Fred W. Ellersick Prize, 2017 IEEE Systems Journal Best Paper Award, and 2018 IEEE CSIM Best Journal Paper Award. He is a Fellow of the Royal Society of Canada, Canadian Academy of Engineering, and Engineering Institute of Canada.

\end{IEEEbiography}

\EOD

\end{document}